\documentclass[useAMS,onecolumn,usenatbib]{mn2e}
\usepackage{graphicx}
\usepackage{times}
\newcommand{\prt}{\partial}

\title[Influence of flow geometry]
{The role of flow geometry in influencing the stability criteria for low angular momentum axisymmetric
black hole accretion}
\author[Nag et al.]
{Sankhasubhra Nag,$^{1}$\thanks{sankhasubhra\_nag@yahoo.co.in}
Swagata Acharya,$^{2}$\thanks{acharya.swagata@gmail.com, Present address:Department of Physics and Metereology, IIT Kharagpur,
Kharagpur, India}
Arnab K. Ray$^{3}$\thanks{arnab.kumar@jiet.ac.in}
and Tapas K. Das$^{4,5}$\thanks{tapas@mri.ernet.in}\\
$^{1}$Department of Physics, Sarojini Naidu College for Women, 
30, Jessore Road, Kolkata 700028, India\\
$^{2}$Department of Physics,
Ramakrishna Mission Vidyamandira,
Belur Math, Kolkata, India \\
$^{3}$Department of Physics, Jaypee University of Engineering and
Technology, A-B Road, Raghogarh, Guna 473226, Madhya Pradesh, India\\
$^{4}$Harish--Chandra Research Institute, Chhatnag Road, Jhunsi,
Allahabad 211019, India\\
$^{5}$Academia Sinica Institute of Astronomy and Astrophysics, No. 1, Roosevelt Rd, Sec. 4, 
Taipei 10617, Taiwan.}

\begin{document}

%\date{}

%\pagerange{\pageref{}--\pageref{}} \pubyear{}

\maketitle

\label{firstpage}

\begin{abstract}
Using mathematical formalism borrowed from dynamical systems theory,
a complete analytical investigation of the critical behaviour of the stationary
flow configuration for the low angular momentum axisymmetric black hole
accretion provides valuable insights about the nature of the phase trajectories
corresponding to the transonic accretion in the steady state, without taking
recourse to the explicit numerical solution commonly performed in the literature
to study the multi-transonic black hole accretion disc and related astrophysical
phenomena. Investigation of the accretion flow around a non rotating black hole
under the influence of various pseudo-Schwarzschild potentials and forming
different geometric configurations of the flow structure manifests that the general
profile of the parameter space divisions describing the multi-critical accretion
is roughly equivalent for various flow geometries. However, a mere variation of
the polytropic index of the flow cannot map a critical solution from one flow geometry
to the another, since the numerical domain of the parameter space responsible to
produce multi-critical accretion does not undergo a continuous transformation in
multi-dimensional parameter space. The stationary configuration used to demonstrate
the aforementioned findings is shown to be stable under linear perturbation for all
kind of flow geometries, black hole potentials, and the corresponding equations of
state used to obtain the critical transonic solutions. Finally, the structure of the
acoustic metric corresponding to the propagation of the linear perturbation studied
are discussed for various flow geometries used.
\end{abstract}


\section{Introduction}
\label{sec1}
Astrophysical blackholes manifest their 
presence only gravitationally. No spectral information 
can directly be obtained from these candidates because of the presence 
of the event horizon. One can only, therefore, rely on accretion 
processes to understand their observational signatures
\citep{pri81,kato-book,fkr02}. At large distances
from the accretor, black hole accretion is usually subsonic. The inner 
boundary condition imposed by the event horizon is determined by the 
requirement that the flow will
be of a supersonic nature very close to the 
accretor. Black hole accretion, thus, usually demonstrates 
transonic behaviour in general. 

Such physical transonic accretion solutions can mathematically be realized 
as critical solutions on the phase portraits of the local radial 
Mach number and the radial distance measured from the event 
horizon 
\citep{rb02,ap03,ray03a,ray03b,rbcqg05a,rbcqg05b,crd06,rbcqg06,rbcqg07a,br07,rb07,gkrd07,jkb09}. 
To maintain physical transonicty such critical 
points will perforce have to be saddle points, which will enable a
solution to pass through themselves. In this connection, a ``multi-critical" 
flow refers to the category of the accretion configuration which can have 
more than one critical points accessible to the flow solution. For low 
angular momentum axisymmetric black hole accretion, it may so happen that 
the critical features are exhibited 
more than once in the phase portrait of a stationary solution 
describing such flows, and accretion consequently becomes multi-critical
\citep{lt80,az81,boz-pac,boz1,fuk83,fuk87,fuk04,fuk04a,lu85,lu86,bmc86,ak89,abram-chak,
ky94,yk95,caditz-tsuruta,das02,dpm03,bdw04,abd06,dbd07,das-czerny}. 

In reality, such weakly rotating sub-Keplerian flows are indeed exhibited in
various physical situations, such as detached binary systems
fed by accretion from OB stellar winds (\cite{ila-shu,liang-nolan}),
semi-detached low-mass non-magnetic binaries (~\cite{bisikalo}),
and super-massive black holes fed
by accretion from slowly rotating central stellar clusters (\cite{ila,ho} and references therein).
Even for a standard Keplerian
accretion disc, turbulence may produce such low angular momentum flows
(see, e.g.,~\cite{igu},
and references therein).

All of the aforementioned multi-critical flow dynamics are important  
in the astrophysical context. Such multi-critical behaviour allows 
the formation of standing shocks in low angular 
momentum axisymmetric black hole accretion
~\citep{fuk83,fuk87,fuk04,fuk04a,c89,ky94,yk95,caditz-tsuruta,
fukumara-suruta,takahashi,das02,dpm03,abd06,dbd07,lyyy97,lugu,nf89,nagyam08,
nakayama,nagakura,toth,das-czerny}. 
Standing shocks in rotating astrophysical accretion potentially 
provide an important and efficient mechanism for conversion 
of a significant amount of the gravitational energy into radiation
by randomizing the directed infall motion of the accreting fluid. Shocks play 
an important role in governing the overall dynamical and radiative processes 
taking place in astrophysical fluid flows around black holes. 

Originally at a large distance, subsonic accretion encounters the outermost 
saddle type critical point and becomes supersonic. Subjected to the 
appropriate perturbative environment, such a supersonic flow encounters a 
shock and becomes subsonic again. The resulting flow has to pass through 
another saddle type critical point to meet the inner boundary condition as 
imposed by the event horizon. For accretion onto a black hole, the presence 
of at least two saddle type critical points is, therefore, a necessary 
(but not sufficient) condition for the shock formation. So multi-critical 
flow behaviour plays a crucial role in studying the physics 
of shock formation and related astrophysical phenomena.

To understand the phase-space behaviour of low angular momentum shocked
multi-transonic accretion, one usually constructs the corresponding 
autonomous dynamical systems analogue, and then identifies the saddle type critical 
points of the phase trajectories of the flow. Next,
the global understanding of the flow topologies are performed which 
necessitates a complete numerical investigation 
of the non-linear stationary equations describing the velocity phase space 
behaviour of the flow. 

It is, however, still possible to semi-quantitatively realize the global
behaviour of the transonic solution without taking resort to numerical
techniques. Getting equipped with the mathematical formalisms of the 
general dynamical systems approach, it has recently been possible
to conceive a clear analytical conception of some of the global 
behaviours of the flow by analyzing the local features of the 
critical points \citep{crd06,mrd07,gkrd07}.

It is important to note that along with understanding the critical 
point behaviour of the stationary accretion solution, it is also necessary to 
ensure that such stationary configuration are stable. This can be accomplished 
by studying the time evolution of a linear acoustic-like perturbation 
(around the stationary configuration) in the full time-dependent 
flow equations. Considering a hydrostatically balanced flow in 
vertical equilibrium, it has recently been observed that for 
accretion onto a non rotating black hole under the influence of 
various pseudo-Schwarzschild black hole potentials, the characteristic 
features of the time development of the aforementioned perturbation 
ensures the stability of the stationary configuration
\citep{crd06}. The result obtained in this way was shown to be 
independent of the choice of the black hole potentials used to study 
accretion flow around a non rotating black hole. 

This work \citep{crd06}, however, was performed for a particular 
type of flow geometry -- hydrostatically balanced flow 
under the vertical equilibrium. Nevertheless, accretion processes onto
astrophysical black holes are also studied for two other different 
flow geometries -- flow with constant disc height, and flow 
under the conical equilibrium \citep{lt80,az81,bl87,lyyy97,bhyasan01,gf03}(see section 2 for further detail
about these two disc models). Those two flow geometries are
relatively simpler to handle (in comparison to the flow configuration 
under the vertical equilibrium) without compromising the essential 
physics involved in the multi-transonic black hole 
accretion phenomena. In addition, these two flows are 
appropriate to study the low angular momentum inviscid flow 
configuration as well. It is thus instructive to investigate 
whether the stationary configurations remain stable (under 
perturbation) in these flow geometries as well. In other words, one needs 
to realize whether the stationary critical solutions are stable irrespective 
of the nature of the space time (choice of the black hole potential) 
as well as the flow geometry (structure of the accretion disc). 

This is precise objective of this work.  
The stationary and the time-dependent low angular momentum 
axisymmetric accretion around a Schwarzschild black hole, 
under the influence of 
a generalized pseudo-Newtonian black-hole potential in  
different flow geometries, have been analyzed. The stationary solutions 
have been considered to 
investigate their critical point behaviour, and to categorize systematically  
the nature of the critical points which appear in such flows. This is followed
by a perturbative study of the  
full time-dependent flow, to follow the evolution of the perturbation and make
predictions about the stability of the stationary configuration.
Finally, observations have also been made about the nature of the 
acoustic metric embedded inside the flow.

\section{The Equations of the flow and its fixed points}
\label{sec2}

When considering a  rotating, axisymmetric, inviscid steady 
flow, the two most pertinent equations are the ones determining the 
drift in the radial direction (essentially Euler's equation),
\begin{equation}
\label{euler}
v \frac{\mathrm{d}v}{\mathrm{d}r} 
+ \frac{1}{\rho}\frac{\mathrm{d}P}{\mathrm{d}r} 
+ \phi^{\prime}(r) - \frac{\lambda^2}{r^3} = 0 
\end{equation}
and the equation of continuity,
\begin{equation}
\label{con}
\frac{\mathrm{d}}{\mathrm{d}r}\left(\rho vrH \right) = 0 \,,
\end{equation}
in which, $\phi(r)$ is the generalised pseudo-Newtonian potential
driving the flow (with the prime denoting a spatial derivative), 
$\lambda$ is the conserved angular momentum of the flow, $P$ is 
the pressure of the flowing gas and $H \equiv H(r)$ is the local 
thickness of the disc, respectively. The two foregoing equations 
give the steady continuum distribution of the velocity field, $v(r)$,
and the density field, $\rho (r)$. But to close the two equations 
it will also be necessary to prescribe the functional dependences 
of both $P$ and $H$
on $v$ and $\rho$, which, in the steady state regime, will imply
an ultimate dependence on $r$. 

Following this requirement, 
the pressure, $P$, is first prescribed by an equation of state for the 
flow~\citep{sc39}. 
As a general polytropic it is given as $P=K \rho^{\gamma}$,
while for an isothermal flow the pressure is given by
$P= \rho {\kappa}T/\mu m_{\mathrm{H}}$, in all of which, 
$K$ is a measure of the entropy in the flow, $\gamma$ is the 
polytropic exponent, $\kappa$ is Boltzmann's constant, $T$ is the 
constant temperature, $m_{\mathrm{H}}$ is the mass of a hydrogen
atom and $\mu$ is the reduced mass, respectively. 

In fixing the function, $H$, one needs to look at the relevant 
vertical geometry of the disc system. This can vary in many ways,
with different degrees of complexity~\citep{bhyasan01}. In the 
simplest case one could treat $H$ to be just a constant, i.e. the 
disc is of uniform thickness. In the case of the conical 
flow~\citep{az81} one
prescribes, $H = Dr$, where $D$ is a simple constant of proportionality. 
While these two cases could be viewed as giving an explicit dependence
of $H$ on $r$,  
another well-invoked, but much more complicated prescription 
in accretion literature is that of the 
disc with the condition of hydrostatic equilibrium imposed in the
vertical direction~\citep{mkfo84,fkr02}. In this particular instance, 
the function $H$ in Eq.~(\ref{con}) will be determined according to the 
way $P$ has been prescribed~\citep{fkr02}. In all of these cases, 
however, it is a common practice to standardise transonicity in the 
flow by scaling its bulk velocity  
with the help of the local speed of sound, which is given as 
$c_{\mathrm{s}} = (\partial P/\partial \rho)^{1/2}$. 
With all these 
analytical requirements stipulated clearly, in what follows, 
the equilibrium properties of the flow will be studied 
for the three different kinds of vertical disc geometry mentioned
above, under both polytropic and isothermal prescriptions for the
equation of state. 

\subsection{Polytropic flows}

With the polytropic relation specified for $P$, it is a 
straightforward exercise to set down in terms of the speed of 
sound, $c_{\mathrm{s}}$, the first integral of Eq.~(\ref{euler}) as, 
\begin{equation}
\label{eupol1st}
\frac{v^2}{2} + n c_{\mathrm{s}}^2 + \phi (r) 
+ \frac{\lambda^2}{2 r^2} = \mathcal{E} 
\end{equation}
in which $n=(\gamma -1)^{-1}$ and the integration constant 
$\mathcal{E}$ is the Bernoulli constant. Before moving on to 
find the first integral of Eq.~(\ref{con}) it should be important
to obtain the functional form of $H$. In simple cases of the 
vertical disc geometry, $H$ usually becomes an explicitly defined
function of the radial distance, $H(r)$ --- either constant disc
height, or a conical profile with a linear dependence, $H = Dr$.
Making a note of this fact, the first
integral of Eq.~(\ref{con}) could be obtained as  
\begin{equation}
\label{conpolaitch}
c_{\mathrm{s}}^{4n}v^2 r^2 H^2  
= \frac{\dot{\mathcal{M}}^2}{4 \pi^2} \,,  
\end{equation}
where $\dot{\mathcal{M}} = (\gamma K)^n \dot{m}$~\citep{az81} 
with $\dot{m}$, an integration constant itself, being physically 
the matter flow rate. 
  
To obtain the critical points of the flow, it should be necessary
first to differentiate both Eqs.~(\ref{eupol1st}) and (\ref{conpolaitch}),
and then, on combining the two resulting expressions, to arrive at
\begin{equation}
\label{dvdrpolaitch}
\left(v^2 - c_{\mathrm{s}}^2 \right)
\frac{\mathrm{d}}{\mathrm{d}r}(v^2) = \frac{2 v^2}{r}
\left[ \frac{\lambda^2}{r^2} - r \phi^{\prime} 
+ c_{\mathrm{s}}^2 \left(1 + r \frac{H^{\prime}}{H}
\right) \right ] \,,
\end{equation}
in which $H^{\prime}$ implies ${\mathrm{d}}H/{\mathrm{d}}r$. 
The critical points of the flow
will be given by the condition that the entire right hand side of
Eq.~(\ref{dvdrpolaitch}) will vanish along with the coefficient of
${\mathrm{d}}(v^2)/{\mathrm{d}r}$. Explicitly written down, and
following some rearrangement of terms, this will give the two
critical point conditions as,
\begin{equation}
\label{critconpolaitch}
%v_{\mathrm{c}}^2 &=& \beta^2 c_{\mathrm{sc}}^2 \nonumber \\
v_{\mathrm{c}}^2 = c_{\mathrm{sc}}^2
= \left[r_{\mathrm{c}} \phi^{\prime}(r_{\mathrm{c}}) 
- \frac{\lambda^2}{r_{\mathrm{c}}^2} \right] 
\left[1 + r_{\mathrm{c}} 
\frac{H^{\prime}(r_{\mathrm{c}})}{H(r_{\mathrm{c}})}\right]^{-1} \,,
\end{equation}
with the subscript ${\mathrm{c}}$ labelling critical point values.
To fix the critical point 
coordinates, $v_{\mathrm{c}}$ and $r_{\mathrm{c}}$,
in terms of the system constants, one would
have to make use of the conditions given by 
Eqs.~(\ref{critconpolaitch}) along
with Eq.~(\ref{eupol1st}), to obtain
\begin{equation}
\label{efixcritaitch}
\frac{1}{2} 
\left(\frac{\gamma +1}{\gamma -1}\right)
\left[r_{\mathrm{c}} \phi^{\prime}(r_{\mathrm{c}})
- \frac{\lambda^2}{r_{\mathrm{c}}^2} \right] 
\left[1 + r_{\mathrm{c}} 
\frac{H^{\prime}(r_{\mathrm{c}})}{H(r_{\mathrm{c}})}\right]^{-1}
+ \phi (r_{\mathrm{c}}) 
+ \frac{\lambda^2}{2 r_{\mathrm{c}}^2} = \mathcal{E} \,,
\end{equation}
from which it is easy to see that solutions of $r_{\mathrm{c}}$ may
be obtained in terms of $\lambda$ and $\mathcal{E}$ only, i.e.
$r_{\mathrm{c}}=f_1(\lambda, \mathcal{E})$.
Alternatively, $r_{\mathrm{c}}$
could be fixed in terms of $\lambda$ and $\dot{\mathcal{M}}$. By making
use of the critical point conditions in Eq.~(\ref{conpolaitch}) one could
write
\begin{equation}
\label{dotmfixaitch}
4 \pi^2 r_{\mathrm{c}}^2 H^2(r_{\mathrm{c}})
\left\{
\left[r_{\mathrm{c}} \phi^{\prime}(r_{\mathrm{c}})
- \frac{\lambda^2}{r_{\mathrm{c}}^2}\right]
\left[1 + r_{\mathrm{c}} 
\frac{H^{\prime}(r_{\mathrm{c}})}{H(r_{\mathrm{c}})}\right]^{-1}
\right\}^{2n +1} = {\dot{\mathcal{M}}}^2 \,,  
\end{equation}
with the obvious implication being that the dependence of $r_{\mathrm{c}}$
will be given as $r_{\mathrm{c}}= f_2(\lambda, \dot{\mathcal{M}})$.
Comparing these two alternative means of fixing $r_{\mathrm{c}}$, the
next logical step would be to say that for the fixed points, and for
the solutions passing through them, it should suffice to specify either
$\mathcal{E}$ or $\dot{\mathcal{M}}$~\citep{skc90}.

For the two relatively simple cases of disc geometry, i.e. constant $H$, 
and $H=Dr$, what Eq.~(\ref{efixcritaitch}) delivers are,  
\begin{equation}
\label{aitchconst}
\frac{1}{2} 
\left(\frac{\gamma +1}{\gamma -1}\right)
\left[r_{\mathrm{c}} \phi^{\prime}(r_{\mathrm{c}})
- \frac{\lambda^2}{r_{\mathrm{c}}^2} \right] 
+ \phi (r_{\mathrm{c}}) 
+ \frac{\lambda^2}{2 r_{\mathrm{c}}^2} = \mathcal{E} 
\end{equation}
and 
\begin{equation}
\label{aitchlin}
\frac{1}{4} 
\left(\frac{\gamma +1}{\gamma -1}\right)
\left[r_{\mathrm{c}} \phi^{\prime}(r_{\mathrm{c}})
- \frac{\lambda^2}{r_{\mathrm{c}}^2} \right] 
+ \phi (r_{\mathrm{c}}) 
+ \frac{\lambda^2}{2 r_{\mathrm{c}}^2} = \mathcal{E} \,,
\end{equation}
respectively. 

All the results obtained so far, can be compared with the case in
which the vertical disc geometry is determined by the condition of 
hydrostatic equilibrium in the vertical direction. This requirement
will deliver the functional form of $H$ as,  
\begin{equation}
\label{aitchpol}
H = c_{\mathrm{s}}\left(\frac{r}{\gamma \phi^{\prime}}\right)^{1/2} \,,
\end{equation}
which evidently shows that $H$ is no more an explicit function of $r$.
Now with the help of this form of $H$, and following the mathematical
procedure outlined so far, 
the critical point coordinates could be fixed in terms of 
the system parameters, $\mathcal{E}$ and ${\dot{\mathcal{M}}}^2$, 
by the relations, 
\begin{equation}
\label{efixcrit}
\frac{2 \gamma}{\gamma -1} 
\left[r_{\mathrm{c}} \phi^{\prime}(r_{\mathrm{c}})
- \frac{\lambda^2}{r_{\mathrm{c}}^2} \right] \left[3 - r_{\mathrm{c}}
\frac{\phi^{\prime \prime}(r_{\mathrm{c}})}{\phi^{\prime}(r_{\mathrm{c}})}
\right]^{-1} + \phi (r_{\mathrm{c}}) 
+ \frac{\lambda^2}{2 r_{\mathrm{c}}^2} = \mathcal{E} 
\end{equation}
and
\begin{equation}
\label{dotmfix}
\frac{4 \pi^2 \beta^2 r_{\mathrm{c}}^3}
{\gamma\phi^{\prime}(r_{\mathrm{c}})}
\left\{\frac{2}{\beta^2}
\left[r_{\mathrm{c}} \phi^{\prime}(r_{\mathrm{c}})
- \frac{\lambda^2}{r_{\mathrm{c}}^2} \right] \left[3 - r_{\mathrm{c}}
\frac{\phi^{\prime \prime}(r_{\mathrm{c}})}{\phi^{\prime}(r_{\mathrm{c}})}
\right]^{-1}\right\}^{2(n +1)} = {\dot{\mathcal{M}}}^2 \,, 
\end{equation}
respectively. A detailed presentation of all
these results (pertaining to the specific case of the disc balanced by
hydrostatic equilibrium in the vertical direction) is to be found in 
a work done by~\citet{crd06}.

\subsection{Isothermal flows}

For isothermal flows, the full mathematical treatment is actually 
much simpler. Here one has to go back to Eq.~(\ref{euler}) and use
the linear dependence between $P$ and $\rho$ as the appropriate equation
of state. On doing so, the first integral of Eq.~(\ref{euler}) is given as
\begin{equation}
\label{euiso1st}
\frac{v^2}{2} + c_{\mathrm s}^2 \ln \rho + \phi (r)
+ \frac{\lambda^2}{2 r^2} = \mathcal{C}
\end{equation}
with $\mathcal{C}$ being a constant of integration. For flow solutions
which specifically decay out to zero at very large distances, the constant 
$\mathcal{C}$ can be determined in terms of the ``ambient conditions" as
$\mathcal{C} = c_{\mathrm s}^2 \ln \rho_\infty$. For the disc of 
constant thickness, or the disc with a conical flow, one is easily 
able to obtain an expression that is identical to Eq.~(\ref{dvdrpolaitch}),
from which the critical point conditions emerge exactly in the 
same form as what has
been shown in Eq.~(\ref{critconpolaitch}). 
However, major point of difference lies in the fact that in an 
isothermal system, the speed of sound, $c_{\mathrm{s}}$, is
globally a constant, and so having arrived at the critical point
conditions, it should be easy to see that $r_{\mathrm{c}}$ and
$v_{\mathrm{c}}$ have already been fixed in terms of a global
constant of the system~\citep{crd06}. 
The speed of sound can further be written
in terms of the temperature of the system as
$c_{\mathrm{s}} = \Theta T^{1/2}$, where
$\Theta = (\kappa/\mu m_{\mathrm H})^{1/2}$, and, therefore,
it should be entirely possible to give a functional dependence for
$r_{\mathrm{c}}$, as $r_{\mathrm{c}} = f_3(\lambda, T)$.
For the two cases of $H$ being a constant, and $H=Dr$, once again
the two respective relations for $r_{\mathrm{c}}$ are
\begin{equation}
\label{aitchiso1}
r_{\mathrm{c}} \phi^{\prime}(r_{\mathrm{c}}) 
- \frac{\lambda^2}{r_{\mathrm{c}}^2} 
= c_{\mathrm{s}}^2 \,, 
\end{equation}
and
\begin{equation}
\label{aitchiso2}
r_{\mathrm{c}} \phi^{\prime}(r_{\mathrm{c}}) 
- \frac{\lambda^2}{r_{\mathrm{c}}^2} 
= 2 c_{\mathrm{s}}^2 \,,
\end{equation}
with the subscript ${\mathrm{c}}$ labelling critical point values,
as usual.

To obtain similar results for the disc in vertical hydrostatic 
equilibrium, it is necessary to go back to Eq.~(\ref{aitchpol}), 
with the restriction that $\gamma =1$, and $c_{\mathrm{s}}$ is 
a global constant. Then it becomes a simple exercise to fix the 
critical point coordinates, by deriving the expression, 
\begin{equation}
\label{critconiso}
%v_{\mathrm{c}}^2 &=& \beta^2 c_{\mathrm{sc}}^2 \nonumber \\
v_{\mathrm{c}}^2 = c_{\mathrm{s}}^2
= 2\left[r_{\mathrm{c}} \phi^{\prime}(r_{\mathrm{c}})
- \frac{\lambda^2}{r_{\mathrm{c}}^2} \right] \left[3 - r_{\mathrm{c}}
\frac{\phi^{\prime \prime}(r_{\mathrm{c}})}{\phi^{\prime}(r_{\mathrm{c}})}
\right]^{-1} \,.
\end{equation}
A detailed derivation of this particular result can once again be
found in the work of~\citet{crd06}. 

\section{Nature of the fixed points : A dynamical systems study}
\label{sec3}

The equations governing the flow in an accreting system are in 
general first-order non-linear differential equations. There is
no standard prescription for a rigorous mathematical analysis of
these equations. Therefore, for any understanding of the behaviour
of the flow solutions, a numerical integration is in most cases 
the only recourse. On the other hand, an alternative approach 
could be made to this question, if the governing equations are
set up to form a standard first-order dynamical system~\citep{js99}. 
This is a very usual practice in general fluid dynamical 
studies~\citep{bdp93}, 
and short of carrying out any numerical integration, this approach
allows for gaining physical insight into the behaviour of the 
flows to a surprising extent. As a first step towards this end,
for the stationary polytropic flow, as given by Eq.~(\ref{dvdrpolaitch}),
it should be necessary to parametrise this equation and set up 
a coupled autonomous first-order dynamical system as~\citep{js99} 
\begin{eqnarray}
\label{dynsysaitch}
\frac{\mathrm{d}}{\mathrm{d}\tau}(v^2)&=& 2v^2 \left[  
\frac{\lambda^2}{r^2} - r \phi^{\prime} 
+ c_{\mathrm{s}}^2 \left(1 + r \frac{H^{\prime}}{H} \right) 
\right] \nonumber \\
\frac{\mathrm{d}r}{\mathrm{d} \tau}&=& r \left(v^2 - 
c_{\mathrm{s}}^2 \right) \,,
\end{eqnarray}
in which $\tau$ is an arbitrary mathematical parameter. With respect
to accretion studies in particular, this kind of parametrisation
has been reported 
before~\citep{rb02,ap03,crd06,mrd07,gkrd07,bbdr09}. 

The critical points have been fixed in terms of the flow
constants. About these fixed point values, upon using a perturbation
prescription of the kind
$v^2 = v_{\mathrm{c}}^2 + \delta v^2$, $c_{\mathrm{s}}^2 = 
c_{\mathrm{sc}}^2 + \delta c_{\mathrm{s}}^2$ and
$r = r_{\mathrm{c}} + \delta r$, one could derive a set of two
autonomous first-order linear differential equations in the
$\delta r$ --- $\delta v^2$ plane, with $\delta c_{\mathrm{s}}^2$ 
having to be first expressed in terms of $\delta r$ and $\delta v^2$,
with the help of Eq.~(\ref{conpolaitch}) --- the continuity equation
--- as
\begin{equation}
\label{varsoundaitch}
\frac{\delta c_{\mathrm{s}}^2}{c_{\mathrm{sc}}^2}=-\left(\gamma -1\right)
\left\{ \frac{\delta v^2}{2v_{\mathrm{c}}^2} 
+ \left[1 + r_{\mathrm{c}} 
\frac{H^{\prime}(r_{\mathrm{c}})}{H(r_{\mathrm{c}})}\right]
\frac{\delta r}{r_{\mathrm{c}}}
\right\} \,.
\end{equation}
The resulting coupled set of linear equations in $\delta r$ and
$\delta v^2$ will follow simply as 
\begin{eqnarray}
\label{lindynsysaitch}
\frac{\mathrm{d}}{\mathrm{d}\tau}
(\delta v^2) &=& -\left(\gamma -1\right)
\left[1 + r_{\mathrm{c}} 
\frac{H^{\prime}(r_{\mathrm{c}})}{H(r_{\mathrm{c}})}\right]
c_{\mathrm{sc}}^2 \,\delta v^2 \nonumber \\
& & - 2c_{\mathrm{sc}}^2
\left[\frac{2 \lambda^2}{r_{\mathrm{c}}^3} + 
\phi^{\prime}(r_{\mathrm{c}}) + r_{\mathrm{c}}\phi^{\prime \prime}
(r_{\mathrm{c}}) + \left(\gamma -1\right)
\left\{1 + r_{\mathrm{c}} 
\frac{H^{\prime}(r_{\mathrm{c}})}{H(r_{\mathrm{c}})}\right\}^2
\frac{c_{\mathrm{sc}}^2}{r_{\mathrm{c}}} 
- c_{\mathrm{sc}}^2\left\{\left(\ln H(r_{\mathrm{c}})\right)^{\prime}
+ r_{\mathrm{c}} \left(\ln H(r_{\mathrm{c}})\right)^{\prime \prime}
\right\}\right] \delta r \nonumber \\
\frac{\mathrm{d}}{\mathrm{d}\tau}(\delta r)
&=& \left(\frac{\gamma + 1}{2}\right)r_{\mathrm{c}}\,\delta v^2 
+ \left(\gamma -1\right) \left[1 + r_{\mathrm{c}} 
\frac{H^{\prime}(r_{\mathrm{c}})}{H(r_{\mathrm{c}})}\right]\delta r \,,
\end{eqnarray}
in which a prime implies a derivative with respect to $r$. 
Trying solutions of the kind $\delta v^2 \sim \exp(\Omega \tau)$
and $\delta r \sim \exp(\Omega \tau)$ in Eqs.~(\ref{lindynsysaitch}), will
deliver the eigenvalues $\Omega$ --- growth rates of $\delta v^2$ and
$\delta r$ --- as
\begin{eqnarray}
\label{eigenaitch}
\Omega^2 &=& \left\{\left(\gamma -1\right)
\left[1 + r_{\mathrm{c}} 
\frac{H^{\prime}(r_{\mathrm{c}})}{H(r_{\mathrm{c}})}\right]
c_{\mathrm{sc}}^2 \right\}^2 \nonumber \\
& & - \left(\gamma +1\right)r_{\mathrm{c}}
c_{\mathrm{sc}}^2
\left[\frac{2 \lambda^2}{r_{\mathrm{c}}^3} + 
\phi^{\prime}(r_{\mathrm{c}}) + r_{\mathrm{c}}\phi^{\prime \prime}
(r_{\mathrm{c}}) + \left(\gamma -1\right)
\left\{1 + r_{\mathrm{c}} 
\frac{H^{\prime}(r_{\mathrm{c}})}{H(r_{\mathrm{c}})}\right\}^2
\frac{c_{\mathrm{sc}}^2}{r_{\mathrm{c}}} 
- c_{\mathrm{sc}}^2\left\{\left(\ln H(r_{\mathrm{c}})\right)^{\prime}
+ r_{\mathrm{c}} \left(\ln H(r_{\mathrm{c}})\right)^{\prime \prime}
\right\}\right] \,. 
\end{eqnarray}
For the specific cases of the simple vertical geometries, i.e. $H$ 
is a constant, and $H=Dr$, the foregoing expression reduces to 
\begin{equation}
\label{eigenaitch1}
\Omega^2 = \left(\gamma -1\right)^2
c_{\mathrm{sc}}^4 
- \left(\gamma +1\right)r_{\mathrm{c}}
c_{\mathrm{sc}}^2
\left[\frac{2 \lambda^2}{r_{\mathrm{c}}^3} + 
\phi^{\prime}(r_{\mathrm{c}}) + r_{\mathrm{c}}\phi^{\prime \prime}
(r_{\mathrm{c}}) + \left(\gamma -1\right)
\frac{c_{\mathrm{sc}}^2}{r_{\mathrm{c}}} 
\right] 
\end{equation}
and
\begin{equation}
\label{eigenaitch2}
\Omega^2 = 4\left(\gamma -1\right)^2
c_{\mathrm{sc}}^4 
- \left(\gamma +1\right)r_{\mathrm{c}}
c_{\mathrm{sc}}^2
\left[\frac{2 \lambda^2}{r_{\mathrm{c}}^3} + 
\phi^{\prime}(r_{\mathrm{c}}) + r_{\mathrm{c}}\phi^{\prime \prime}
(r_{\mathrm{c}}) + 4 \left(\gamma -1\right)
\frac{c_{\mathrm{sc}}^2}{r_{\mathrm{c}}} 
\right] \,, 
\end{equation}
respectively. 

A similar treatment might also be extended to the case of the
disc in vertical hydrostatic equilibrium. 
Following a perturbative treatment about the fixed point coordinates,  
the eigenvalues, $\Omega$, could be obtained in this instance as 
\begin{equation}
\label{eigen}
\Omega^2 = \frac{4 r_{\mathrm{c}}  
\phi^{\prime}(r_{\mathrm{c}})c_{\mathrm{sc}}^2}{(\gamma + 1)^2} 
\left\{ \left[ \left(\gamma - 1 \right)
{\mathcal A} - 2 \gamma \left(4 + {\mathcal A} \right) + 2 \gamma 
{\mathcal{B}}\left(1 + \frac{3}{\mathcal A}\right) \right] 
- \frac{\lambda^2}{\lambda_{\mathrm K}^2(r_{\mathrm{c}})} 
\left[4 \gamma + \left(\gamma - 1 \right){\mathcal A} + 2 \gamma 
{\mathcal{B}}\left(1 + \frac{3}{\mathcal A}\right) \right]
\right\} \,,
\end{equation}
where $\lambda_{\mathrm K}^2(r) = r^3 \phi^{\prime}(r)$, and 
with 
\begin{displaymath}
\label{coeffs}
\mathcal{A} = r_{\mathrm{c}}\frac{\phi^{\prime \prime}(r_{\mathrm{c}})}
{\phi^{\prime}(r_{\mathrm{c}})} - 3 \,,  \qquad
\mathcal{B} = 1 + r_{\mathrm{c}}
\frac{\phi^{\prime \prime \prime}(r_{\mathrm{c}})}
{\phi^{\prime \prime}(r_{\mathrm{c}})}
- r_{\mathrm{c}}\frac{\phi^{\prime \prime}(r_{\mathrm{c}})}
{\phi^{\prime}(r_{\mathrm{c}})} \,.
\end{displaymath}
The detailed calculations to arrive at Eq.~(\ref{eigen}) 
could once again be accessed in the work of~\citet{crd06}.  

For isothermal flows, similar expressions for the related eigenvalues 
may likewise be derived, given a particular form of the function, $H(r)$. 
The algebra in this case is much simpler and it is an easy exercise 
to assure oneself that for isothermal flows one just needs to set 
$\gamma = 1$ in 
Eqs.~(\ref{eigenaitch}),~(\ref{eigenaitch1},~(\ref{eigenaitch2}) 
and (\ref{eigen}), to arrive at a corresponding 
relation for $\Omega^2$. However, it should be incorrect to assume 
that in this kind of study, one could always treat isothermal flows 
simply as a special physical case of general polytopic flows. 
For polytropic flows, the position of the fixed points, under a given
form of $\phi(r)$, will be determined by 
Eq.~(\ref{efixcritaitch}) or by Eq.~(\ref{dotmfixaitch}), in the
case of simple vertical disc geometries, and by Eq.~(\ref{efixcrit}) 
or by Eq.~(\ref{dotmfix}), in the case of the disc balanced by 
hydrostatic equilibrium in the vertical direction. 
On the other hand, for isothermal flows the fixed points are 
simply to be determined from the critical point conditions themselves, 
since $c_{\mathrm{s}}$ is globally a constant in 
this case~\citep{crd06}. The resulting difference is by no means trivial.   

Once the position of a critical point, $r_{\mathrm{c}}$, has been 
ascertained, it is then a straightforward task to find the nature 
of that critical point by using $r_{\mathrm{c}}$ in either 
Eq.~(\ref{eigenaitch}) or Eq.~(\ref{eigen}), depending on the 
disc geometry.  
Since it has been discussed in Section~\ref{sec2} that $r_{\mathrm{c}}$
is a function of $\lambda$ and $T$ for isothermal flows, and a function
of $\lambda$ and $\mathcal E$ (or $\dot{\mathcal M}$) for polytropic
flows, it effectively implies that $\Omega^2$ can, in principle, be 
rendered as a function of the flow parameters for either kind of flow. 
A generic conclusion that can be drawn about the critical points from 
the form of $\Omega^2$ in Eqs.~(\ref{eigenaitch}) and (\ref{eigen}), 
is that the only 
admissible critical points will be saddle points and centre-type points. 
For a saddle point, $\Omega^2 > 0$, while for a centre-type point, 
$\Omega^2 < 0$. 

\section{Numerical results} 
\label{sec4}
It is quite evident from previous discussions that the location of critical points and their nature (saddle or centre) 
are easily determined from the roots of the equations (Eq.\ref{aitchconst}-\ref{aitchpol},\ref{aitchiso1}-\ref{critconiso}) 
and the sign of $\Omega^2$. These, in turn, effectively extract the qualitative features of the phase portrait without 
having an explicit plot of it.

In this work, with numerical evidence, these tools are exploited to compare the features of the accretion flow in 
different disc geometries under polytropic as well as isothermal equations of state. Though this formulation of the 
procedure works for any choice of a pseudo-Schwarchild potential, the Paczynski-Witta 
potential \citep{pw80},
\begin{equation}
\label{potens}
\phi_{PW} (r) = - \frac{1}{2 \left(r - 1 \right)} \,,
\end{equation}
has been selected for the presentation of numerical computations within pseudo-Newtonian framework.
 
\begin{figure}
	\centering
		\includegraphics[width=0.4\textwidth]{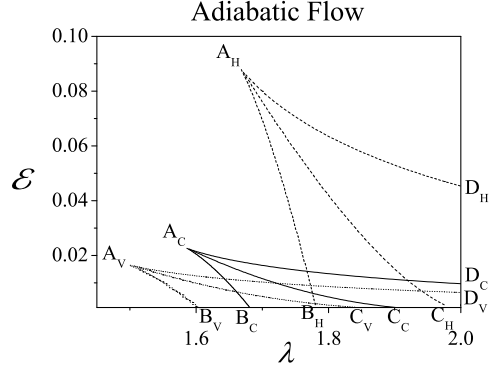}
	\caption{Different regions in parameter space of $\mathcal{E}$ and $\lambda.$ corresponding to number 
and nature of critical points for polytropic flows. The dotted lines are for vertical equilibrium geometry, the solid lines 
are for conical geometry, and the small dashed lines are for constant-height disc. For notations see text.}
	\label{fig:AdiBound}
\end{figure}
\begin{figure}
	\centering
		\includegraphics[width=0.4\textwidth]{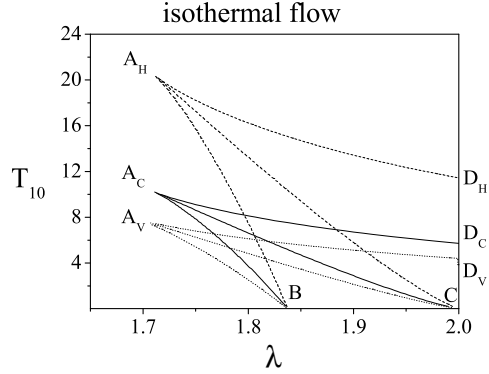}
	\caption{Different regions in parameter space of $T_{10}$ and $\lambda$ corresponding to the number and nature of 
critical points for polytropic flows. $T_{10}$ means temperature of the flow in $10^{10}$$^oK$. Linestyles are similar to 
the previous figure. For other notations see text.}
	\label{fig:IsoBound}
\end{figure}
First of all, Fig.~\ref{fig:AdiBound} shows the variation of the number of critical points for polytropic flows 
with the variation of $\mathcal{E}$ and $\lambda$ for the three disc geometries. For example, the region bounded 
by $ A_H$, $ B_H$ and $ D_H$ depicts the parameter values corresponding to three critical points in the constant energy 
flow solutions, for the constant-height disc geometry, within which for the subregion bounded by $ A_H$, $ B_H$, and $ C_H$ 
the value of $ \dot{\cal{M}}$ at the innermost critical point ($\dot{\cal{M}}_{in}$ henceforth) is higher than that at 
the outermost critical point ($ \dot{\cal{M}}_{out}$). The relative values of $\dot{\cal{M}}$ are just the reverse for 
the other subregion bounded by $ A_H$, $ C_H$ and $ D_H$. On the curve $ A_H C_H$ at all points $ \dot{\cal{M}}$ acquires 
the same value at both the critical points; for all other regions shown in the graph there will exist only single 
critical points. 

For other disc models, similar features are depicted by the symbols with suffixes $C$ (Conical) and $V$ (Vertical 
equilibrium). Hence for all three geometries, there are certain wedge shaped regions which correspond to three critical 
points and outside these regions, the parameter values within the parameter space shown in the figure, generate single 
critical points only. Hence for low values of $\mathcal{E}$ with there will be an interval of values of $\lambda$ for 
which three critical points exist (opening up the the possibility of multi-transonic accretion within the subregion 
where  $\dot{\cal{M}}_{in}>\dot{\cal{M}}_{out}$ under certain other condition (see later)).
In Fig.~\ref{fig:IsoBound} the same thing is shown with the relevant parameters $T$ and $\lambda$. Here also the 
appearance of the wedge shaped region conforms to the possibility of multi-transonic accretion for certain interval 
of values of $\lambda$. But here in the $ c_s^2=\Theta^2 T\rightarrow 0 $ limit, all three equations merge together 
and as a result of it, in the figure the lower points $B$ and $C$ for all models, merge together. Instead 
of $\dot{\cal{M}}$, here $\cal{C}$ is the quantity which makes a difference between subregions $ABC$ and $ACD$ in 
different models. 

Both in polytropic and isothermal cases, the interval of $\lambda$ permitting more than one critical points 
progressively shrinks for higher values of the other parameter (i.e. $\mathcal{E}$ and $T$, respectively) and 
there are certain values of $\mathcal{E}$ or $T$, dependent on the disc geometry, over which there will be only 
a single critical point, whatever be the values of $\lambda$. The wedge shaped regions also widely vary from one 
disc geometry to another. Actually the regions of parametric values allowing multicritical points for various disk 
geometries may be approximately mapped from one another with some suitable scaling of $\gamma$ (\citep{bhyasan01}); 
but if one may map such a parametric region for the polytropic flow into another correspondoing parametric region 
with a choice of $\gamma$ (for the second case) nearly equal to unity, it should not create the impression that 
the first flow in the polytropic condition may at least approximately mimic the second flow in the isothermal condition, 
because the physical conditions and the first integral of motions under the two conditions (polytropic and isothermal) 
are distinctly different.

\begin{figure}[htb]
	\centering
		\includegraphics[width=0.75\textwidth]{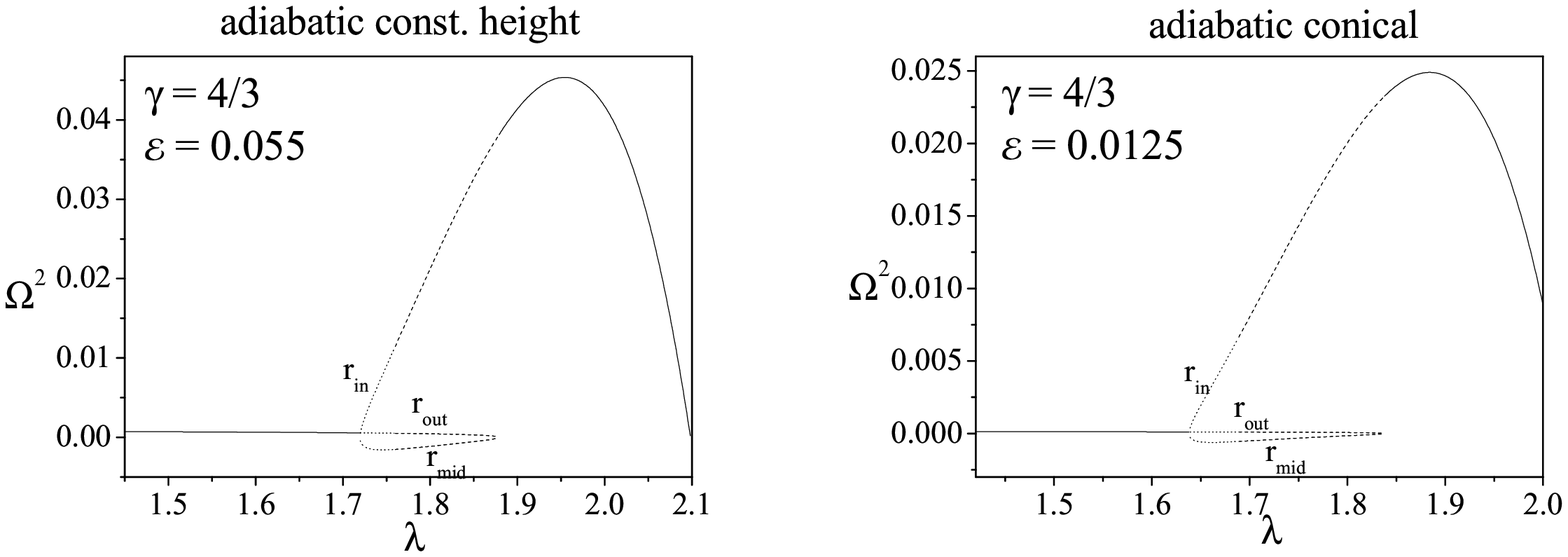}
	\caption{Variation of $\Omega^2$ with $\lambda$ at constant $\cal{E}$ for polytropic flow.}
	\label{fig:omegasqAdi}
\end{figure}
\begin{figure}[htb]
	\centering
		\includegraphics[width=0.75\textwidth]{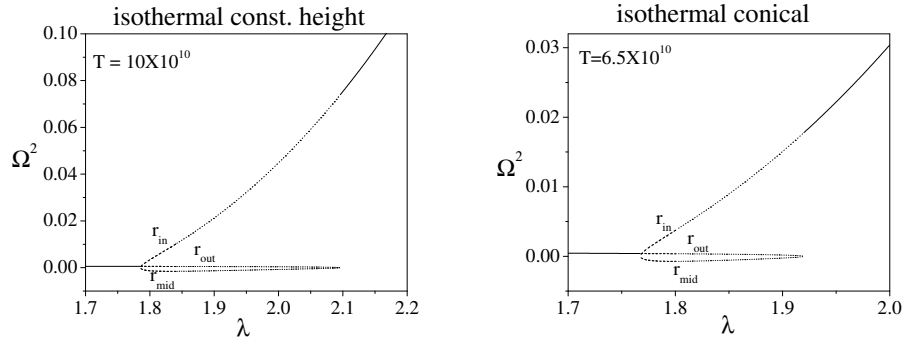}
	\caption{Variation of $\Omega^2$ with $\lambda$ at constant $T$ for isothermal flow.}
	\label{fig:isoOmega}
\end{figure}	
The nature of the critical points will be apparent from the Fig.~\ref{fig:omegasqAdi} and Fig.~\ref{fig:isoOmega}. 
In these plots a positive value of $\Omega^2$ indicates a saddle point while a negative value does the same for 
a centre. A common feature of both the separate plots is that initially there is region of a single saddle point, 
followed by the birth of a centre-type point and another saddle point (saddle-centre type bifurcation); then at 
another higher value of $\lambda$, the centre-type point coalesces with the other saddle point (the outer one) 
and both of them annihilate each other (another saddle-centre type bifurcation, but this time in the opposite 
direction) so that the remaining saddle point (outer) ``survives" above the critical value of $\lambda$. Actually 
in the underlying phase plots (not shown here), among the three critical points, there will always be a pair of 
centre-saddle for which the separatrices  of the saddle form a homoclinic connection around the centre, as long 
as the other constant of flow ($\dot{\cal{M}}$ for polytropic case and $\cal{C}$ for isothermal case) discriminates 
between the two saddle points. The remaining saddle point will have the separatrices connecting the event horizon 
with the infinity. The separatrices of the saddle point with higher value of $\dot{\cal{M}}$ in the polytropic 
case (with lower value of $\cal{C}$ in isothermal case) form the homoclinic connection. Physically this is what 
it has to be, because it is the third critical point (saddle) which will allow the transonic flow solution from 
infinity to the event horizon, and here, following the line of argument in \cite{rb02}, it can be stated that 
the stationary flow solution has to settle on the separatices of a saddle point only. That is why among the 
critical points, only the saddle points are termed as sonic points, and not the centre-type points, although 
for both the types of critical points the flow speed is equal to the sound speed, i.e. Mach number becomes unity.
 
So if the value of $\lambda$ is increased from a sufficiently low value, first there will a span of single 
saddle point (solid line in the figure) then the bifurcation occurs resulting in a pair of centre-type points 
(with intermediate value $r$, denoted by $r_{mid}$) and a new saddle point (location $r_{in}$) so that the 
older saddle point becomes the outermost critical point (location $r_{out}$). Before $\lambda$ attains a 
certain value, 
the relation $\dot{\cal{M}}_{in}>\dot{\cal{M}}_{out}$ (for isothermal case ${\cal{C}}_{in}<{\cal{C}}_{out}$) 
is maintained and for this interval of $\lambda$, the $\Omega^2$ functions for the three critical points are 
plotted with dotted lines. Here the separatrices of the newly formed saddle point encompasses the centre-type
point, forming a homoclinic connection; after exceeding this value of $\lambda$, what happens is that  
$\dot{\cal{M}}_{in}>\dot{\cal{M}}_{out}$, and the homoclinic connection for the inner saddle opens up, 
giving rise to the same sort of homoclinic connection from the outer saddle. Within this phase, the function
of $\Omega^2$ is depicted with small dashed curves. At the end of this domain of $\lambda$, the centre-type 
and the outer saddle point coalesce and ``destroy" each other, after which, only the inner saddle point exists 
with increasing value of $\lambda$ (solid curves).

The scenario in the isothermal case is almost similar. There is an interesting situation along the boundary 
between the two subregions within the wedge shaped region corresponding to multicritical points. For these 
parameter values $\dot{\cal{M}}_{in}=\dot{\cal{M}}_{out}$, and thus two saddle points will be linked by 
a heteroclinic connection of their separatrices; for the $\lambda$ just below this value there is some 
homoclinic connection for the inner saddle point, and just above this value the homoclinic connection 
corresponds to the outer critical points. From the point of view of a general dynamical system along this 
boundary, a type of bifurcation occurs that may be termed as a heteroclinic one.

\begin{figure}
	\centering
		\includegraphics[width=0.45\textwidth]{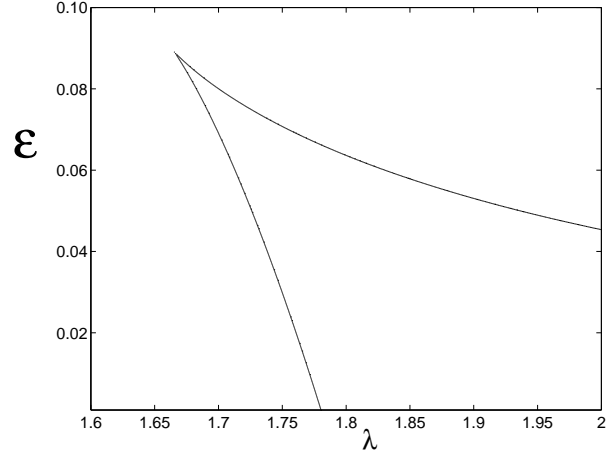}
	\caption{Contour plot of $D$ for $D=0$.}
	\label{fig:contour}
\end{figure}

The boundaries of the region in the parameter space permitting multicritical points, as it has been discussed
already, are associated with saddle-centre bifurcation or merging of a pair of roots of the 
equations (Eq.\ref{aitchconst}-\ref{aitchpol}, \ref{aitchiso1}-\ref{critconiso}) needed for determining 
the critical points under various conditions discussed above. Now all these equations are polynomial equations. 
The discriminant of a general polynomial,
\begin{equation}
	P_n(x)=a_nx^n+a_{n-1}x^{n-1}+\cdots+a_1x+a_0,
\end{equation}
 can be expressed as in terms of its roots, $x_i$'s, as
\begin{equation}
	D=a_n^{n-2}\prod_{i<j}{(x_i-x_j)^2}.   
\end{equation}
The discriminant may be expressed as the determinant of a matrix called 
Sylvester matrix (http://mathworld.wolfram.com/PolynomialDiscriminant.html and other references therein),
\begin{equation}
	S=\left[\begin{array}{lllllllll}
\multicolumn{1}{c}{a_n} & \multicolumn{1}{c}{a_{n-1}} & \multicolumn{1}{c}{a_{n-2}} & \multicolumn{1}{c}{\ldots} & \multicolumn{1}{c}{a_1} & \multicolumn{1}{c}{a_0} & \multicolumn{1}{c}{0\ldots} & \multicolumn{1}{c}{\ldots} & \multicolumn{1}{c}{0} \\ 
\multicolumn{1}{c}{0} & \multicolumn{1}{c}{a_n} & \multicolumn{1}{c}{a_{n-1}} & \multicolumn{1}{c}{a_{n-2}} & \multicolumn{1}{c}{\ldots} & \multicolumn{1}{c}{a_1} & \multicolumn{1}{c}{a_0} & \multicolumn{1}{c}{0\ldots} & \multicolumn{1}{c}{0} \\ 
\multicolumn{1}{c}{\vdots} & \multicolumn{1}{c}{} & \multicolumn{1}{c}{} & \multicolumn{1}{c}{} & \multicolumn{1}{c}{} & \multicolumn{1}{c}{} & \multicolumn{1}{c}{} & \multicolumn{1}{c}{} & \multicolumn{1}{c}{\vdots} \\ 
\multicolumn{1}{c}{0} & \multicolumn{1}{c}{\ldots} & \multicolumn{1}{c}{0} & \multicolumn{1}{c}{a_n} & \multicolumn{1}{c}{a_{n-1}} & \multicolumn{1}{c}{a_{n-2}} & \multicolumn{1}{c}{\ldots} & \multicolumn{1}{c}{a_1} & \multicolumn{1}{c}{a_0} \\ 
\multicolumn{1}{c}{na_n} & \multicolumn{1}{c}{(n-1)a_{n-1}} & \multicolumn{1}{c}{(n-2)a_{n-2}} & \multicolumn{1}{c}{\ldots} & \multicolumn{1}{c}{1a_1} & \multicolumn{1}{c}{0} & \multicolumn{1}{c}{\ldots} & \multicolumn{1}{c}{\ldots} & \multicolumn{1}{c}{0} \\ 
\multicolumn{1}{c}{0} & \multicolumn{1}{c}{na_n} & \multicolumn{1}{c}{(n-1)a_{n-1}} & \multicolumn{1}{c}{(n-2)a_{n-2}} & \multicolumn{1}{c}{\ldots} & \multicolumn{1}{c}{1a_1} & \multicolumn{1}{c}{0} & \multicolumn{1}{c}{\ldots} & \multicolumn{1}{c}{0} \\ 
\multicolumn{1}{c}{\vdots} & \multicolumn{1}{c}{} & \multicolumn{1}{c}{} & \multicolumn{1}{c}{} & \multicolumn{1}{c}{} & \multicolumn{1}{c}{} & \multicolumn{1}{c}{} & \multicolumn{1}{c}{} & \multicolumn{1}{c}{\vdots} \\ 
\multicolumn{1}{c}{0} & \multicolumn{1}{c}{0} & \multicolumn{1}{c}{\ldots} & \multicolumn{1}{c}{0} & \multicolumn{1}{c}{na_n} & \multicolumn{1}{c}{(n-1)a_{n-1}} & \multicolumn{1}{c}{(n-2)a_{n-2}} & \multicolumn{1}{c}{\ldots} & \multicolumn{1}{c}{1a_1} \\ 
\end{array}\right],
\end{equation}
up to a factor.
Hence it is expected that $\det{(S)}$ will be zero on the abovementioned boundaries and actually it is so. 
Here the contour plot of $\det{(S)}=0$ value for the polytropic flow in constant height geometry (i.e. for 
the polynomial in $r_c$ in Eq.\ref{aitchconst}) in $\cal{E}$--$\lambda$ space is shown in Fig.\ref{fig:contour}. 
The curve exactly conforms with the corresponding boundary curve (in small dashed style) in Fig.\ref{fig:AdiBound}, 
drawn on the basis of the numerical method of detailed root finding. So this procedure may be thought of as a 
much easier alternative to find the multicritical parametric values.

Considering now the question of actual accretion process, from Eq.\ref{dvdrpolaitch} it is evident that 
for conical and constant height model the critical points are actual sonic points, though these two are 
not exactly same for the other model.% (Eq.\ref{critconpolaitch}). 
As for all parameter values there will always be one critical (sonic) point with separatrices spanning 
from infinity to the event horizon, there will always exist a transonic solution. But the features of 
the critical points in all the three models opens the possibility of multitransonic accretion too 
within the region with multicritical points under certain conditions. First, there should be some 
transition from the supersonic branch of the non-homoclinic separatrices (trajectory) of the critical 
point (sonic point) with global transonic solution to the subsonic branch of the homoclinic solution 
of the other sonic point at a particular value of $r$. Secondly there should be some quantity which 
will discriminate between the trajectories of these two sonic points. For such a transition one such 
possibility is the occurrence of a shock, which may take place obeying the 
condition $\dot{\cal{M}}_{in}>\dot{\cal{M}}_{out}$ for the polytropic case 
(and ${\cal{C}}_{in}<{\cal{C}}_{out}$ for the isothermal condition). After making the transition, 
the flow will pass through the corresponding inner sonic point. The determination of the exact 
location and the exact parameter values for such a transition really taking place is out of the 
scope of the present paper, but here it is touched upon to indicated that this full analysis 
keeps alive the possibility for such types of multitransonic accretion, irrespective of the disc 
geometries and equations of state. 

\section{Time-dependent stability analysis of stationary solutions}
\label{sec5}

The time-dependent generalisation of the continuity condition for
an axisymmetric pseudo-Schwarzschild disc is given as 
\begin{equation}
\label{surden}
\frac{\prt \Sigma}{\prt t}+ \frac{1}{r} \frac{\prt}{\prt r}
\left( \Sigma vr \right)=0 \,,
\end{equation}
in which the surface density of the  disc, $\Sigma$, is to be
expressed as $\Sigma \cong \rho H$~\citep{fkr02}. From the foregoing
expression, one can, therefore, obtain
\begin{equation}
\label{voldenh}
\frac{\prt \rho}{\prt t}+ \frac{1}{rH} \frac{\prt}{\prt r}
\left( \rho vrH \right)=0 \,.
\end{equation}
Defining a new variable $f = \rho v rH$,
%$f=\rho^{(\gamma +1)/2} v r^{3/2}/\sqrt{\phi^{\prime}}$, 
it is quite
obvious from the form of Eq.~(\ref{surden}) that the stationary value 
of $f$ will be a constant, $f_0$, which can be closely identified with
the matter flux rate. This follows a similar approach to spherically 
symmetric flows made by~\citet{pso80} and~\citet{td92}. For an 
axisymmetric disc, which has no dependence on any angle variable, 
this approach has also been adopted for a flow driven simply by a 
Newtonian potential~\citep{ray03}. The present treatment, of course,
is of a more general nature, the disc flow being driven by a general
pseudo-Newtonian potential, $\phi (r)$ for all disk geometries. In this system, a perturbation 
prescription of the form $v(r,t) = v_0(r) + v^{\prime}(r,t)$ and 
$\rho (r,t) = \rho_0 (r) + \rho^{\prime}(r,t)$, will give, on 
linearising in the primed quantities,  
\begin{equation}
\label{effprimeh}
\frac{f^{\prime}}{f_0} = \frac{\rho^{\prime}}{\rho_0} 
+ \frac{v^{\prime}}{v_0} \,,
\end{equation}
which is 
a relation that connects all the three 
fluctuating quantities, $v^{\prime}$,  $\rho^{\prime}$  
and $f^{\prime}$, with one another (here the subscript $0$ denotes 
stationary background 
values in all the cases). Going back to Eq.~(\ref{voldenh}), 
it becomes possible to connect $\rho^{\prime}$ exclusively to 
$f^{\prime}$, through the relation 
\begin{equation}
\label{flucdenh}
\frac{\prt \rho^{\prime}}{\prt t} + 
\frac{v_0 \rho_0}{f_0} 
\left(\frac{\prt f^{\prime}}{\prt r}\right)=0 \,.
\end{equation}

The case of the disc being balanced vertically under hydrostatic
equilibrium is very different in mathematical terms, 
and has been taken up 
earlier by~\citet{crd06}. Nevertheless, it is worth going back
to it for making some interesting comparisons. 
By using Eq.~(\ref{aitchpol}) and the polytropic 
relation $P = K \rho^{\gamma}$, Eq.~(\ref{surden})
can be rendered as
\begin{equation}
\label{volden}
\frac{\prt}{\prt t} \left[\rho^{(\gamma +1)/2}\right]
+ \frac{\sqrt{\phi^{\prime}}}{r^{3/2}}
\frac{\prt}{\prt r} \left[ \rho^{(\gamma +1)/2} v
\frac{r^{3/2}}{\sqrt{\phi^{\prime}}} \right] =0 \,,
\end{equation}
from which, under a new definition, 
$f=\rho^{(\gamma +1)/2} v r^{3/2}/\sqrt{\phi^{\prime}}$, 
one obtains
\begin{equation}
\label{effprime}
\frac{f^{\prime}}{f_0}=\left( \frac{\gamma +1}{2} \right)
\frac{\rho^{\prime}}
{\rho_0} + \frac{v^{\prime}}{v_0} \,.
\end{equation}
From Eq.~(\ref{volden}), it is also very easy to set down the density
fluctuations, $\rho^{\prime}$, in terms of $f^{\prime}$, as 
\begin{equation}
\label{flucden}
\frac{\prt \rho^{\prime}}{\prt t} + \beta^2
\frac{v_0 \rho_0}{f_0} \left(\frac{\prt f^{\prime}}{\prt r}\right)=0 \,,
\end{equation}
with $\beta^2 = 2(\gamma +1)^{-1}$, as before. This result may be 
compared with Eq.~(\ref{flucdenh}) and the difference noted. 
If, however, one were to study an
isothermal flow balanced by hydrostatic equilibrium in the vertical
direction, then Eq.~(\ref{aitchpol}) would have to be constrained by 
$\gamma =1$ and $c_{\mathrm{s}}$ being constant. Under these
conditions, the expression for density fluctuations in the flow will 
be identical to Eq.~(\ref{flucdenh}), rather than be described by
Eq.~(\ref{flucden}). 

The equation 
for density fluctuations may show variations in terms of a constant 
scaling factor under different vertical disc height geometries, but 
no such thing happens for the velocity fluctuations. 
Combining either Eqs.~(\ref{effprimeh}) and (\ref{flucdenh}), or 
combining Eqs.~(\ref{effprime}) and (\ref{flucden}) will render
the velocity fluctuations as 
\begin{equation}
\label{flucvel}
\frac{\prt v^{\prime}}{\prt t}= \frac{v_0}{f_0}
\left(\frac{\prt f^{\prime}}{\prt t}+{v_0}
\frac{\prt f^{\prime}}{\prt r}\right) \,,
\end{equation}
which, upon a further partial differentiation with respect to time,
will give 
\begin{equation}
\label{flucvelder2}
\frac{{\prt}^2 v^{\prime}}{\prt t^2}=\frac{\prt}{\prt t} \left[
\frac{v_0}{f_0} \left(\frac{\prt f^{\prime}}{\prt t}\right) \right]
+ \frac{\prt}{\prt t} \left[ \frac{v_0^2}{f_0} \left(
\frac{\prt f^{\prime}}{\prt r}\right) \right] \,.
\end{equation}

The time-dependent equation for the radial drift is given as 
\begin{equation}
\label{dyneuler}
\frac{\prt v}{\prt t} + v \frac{\prt v}{\prt r}
+ \frac{1}{\rho} \frac{\prt P}{\prt r} + \phi^{\prime}(r)
-\frac{\lambda^2}{r^3}=0 \,,
\end{equation}
from which the linearised fluctuating part could be
extracted as 
\begin{equation}
\label{fluceuler}
\frac{\prt v^{\prime}}{\prt t}+ \frac{\prt}{\prt r}
\left( v_0 v^{\prime} + c_{\mathrm{s0}}^2 
\frac{\rho^{\prime}}{\rho_0}\right) =0 \,,
\end{equation}
with $c_{\mathrm{s0}}$ being the speed of sound in the steady state.
Differentiating Eq.~(\ref{fluceuler}) partially with respect to $t$,
and making use of either Eq.~(\ref{flucdenh}) or Eq.~(\ref{flucden}), 
along with Eqs.~(\ref{flucvel}) and  
(\ref{flucvelder2}), to substitute for all the first and second-order
derivatives of $v^{\prime}$ and $\rho^{\prime}$, will deliver the result 
\begin{equation}
\label{interm}
\frac{\prt}{\prt t} \left[\frac{v_0}{f_0}
\left( \frac{\prt f^{\prime}}{\prt t}\right)\right]
+ \frac{\prt}{\prt t} \left[\frac{v_0^2}{f_0}
\left( \frac{\prt f^{\prime}}{\prt r}\right)\right]
+ \frac{\prt}{\prt r} \left[\frac{v_0^2}{f_0}
\left( \frac{\prt f^{\prime}}{\prt t}\right)\right]
+ \frac{\prt}{\prt r} \left[\frac{v_0}{f_0}
\left(v_0^2 - \sigma c_{\mathrm{s0}}^2 \right)
\frac{\prt f^{\prime}}{\prt r}\right] = 0 \,,
\end{equation}
in which either $\sigma =1$ or $\sigma = \beta^2$, depending on the 
choice of a particular disc geometry and the equation state applied. 
For isothermal flows, $\sigma =1$, for whatever disc geometry one 
considers --- $H$ is constant or $H=Dr$ or $H$ is as it is described 
by Eq.~(\ref{aitchpol}). The same value of $\sigma$ is also obtained 
for polytropic flows in the first two simple cases of the height
function, $H$. The common feature running through all these cases 
is that $H$ in Eq.~(\ref{voldenh}) does not have any 
dependence on time. It is only when the flow is polytropic and 
the disc height geometry is expressed by Eq.~(\ref{aitchpol}), will
$H$ have a time-dependence, whose ultimate consequence will be that
$\sigma = \beta^2$ in Eq.~(\ref{interm}). 

All the terms in Eq.~(\ref{interm}) can be expediently rendered 
into a compact formulation that looks like
\begin{equation}
\label{compact}
\prt_\mu \left( {\mathrm{f}}^{\mu \nu} \prt_\nu 
f^{\prime}\right) = 0 \,,
\end{equation}
in which the Greek indices are made to run from $0$ to $1$, with 
the identification that $0$ stands for $t$, and $1$ stands for $r$.
An inspection of the terms in the left hand side of Eq.~(\ref{interm})
will then allow for constructing the symmetric matrix
\begin{equation}
\label{matrix}
{\mathrm{f}}^{\mu \nu } = \frac{v_0}{f_0}
\pmatrix
{1 & v_0 \cr 
v_0 & v_0^2 - \sigma c_{\mathrm{s0}}^2} \,.
\end{equation}
Now the d'Alembertian for a scalar in curved space is given in terms
of the metric ${\mathrm{g}}_{\mu \nu}$ by~\citep{vis98}
\begin{equation}
\label{alem}
\Delta \psi \equiv \frac{1}{\sqrt{-\mathrm{g}}}
\prt_\mu \left({\sqrt{-\mathrm{g}}}\, {\mathrm{g}}^{\mu \nu} \prt_\nu
\psi \right) \,,
\end{equation}
with $\mathrm{g}^{\mu \nu}$ being the inverse of the matrix implied
by ${\mathrm{g}}_{\mu \nu}$. Using the equivalence that
${\mathrm{f}}^{\mu \nu } = \sqrt{-\mathrm{g}}\, {\mathrm{g}}^{\mu \nu}$,
and therefore $\mathrm{g} = \det \left({\mathrm{f}}^{\mu \nu }\right)$,
it is immediately possible to set down an effective metric for the 
propagation of an acoustic disturbance as
\begin{equation}
\label{metric}
\mathrm{g}^{\mu \nu}_{\mathrm{eff}} =
\pmatrix
{1 & v_0 \cr
v_0 & v_0^2 - \sigma c_{\mathrm{s0}}^2} \,,
\end{equation}
which can be shown to be entirely identical to the metric of a wave 
equation for a scalar field in curved space-time, obtained through a
somewhat different approach~\citep{vis98}. The inverse effective
metric, $\mathrm{g}_{\mu \nu}^{\mathrm{eff}}$, can be easily derived
by inversion of the matrix given in Eq.~(\ref{metric}), and this will 
give $v_0^2 = \sigma c_{\mathrm{s0}}^2$ as the horizon condition of an
acoustic black hole for inflow solutions~\citep{vis98}. 
From the perspective of the propagation of acoustic waves (carrying
information in any fluid system) in the accretion disc, what can be 
concluded from these arguments is that the disc geometry and the equation
of state act together in determining the speed of information propagation. 
In the case of the disc being supported by hydrostatic equilibrium
in the vertical direction, if the flow is polytropic, then the speed
of information propagation will be less than the speed of sound by 
a factor, $\beta$. So transonicity will not take place exactly when
the bulk flow speed, $v$, becomes equal to the speed of 
sound, $c_{\mathrm{s}}$. In all the other cases, transonicity will,
however, take place when $v = c_{\mathrm{s}}$. 

Finally, a little readjustment of terms in Eq.~(\ref{interm}) will 
give an equation for the perturbation as  
\begin{equation}
\label{tpert}
\frac{{\prt}^2 f^{\prime}}{\prt t^2} +2 \frac{\prt}{\prt r}
\left(v_0 \frac{\prt f^{\prime}}{\prt t} \right) + \frac{1}{v_0}
\frac{\prt}{\prt r}\left[ v_0 \left(v_0^2- 
\sigma c_{\mathrm{s0}}^2 \right)
\frac{\prt f^{\prime}}{\prt r}\right] = 0 \,,
\end{equation}
whose detailed solution has been given in earlier works on inviscid 
axisymmetric flows~\citep{ray03,crd06}. 

\section{Concluding remarks}
\label{sec6}
The primary motivation behind this work, as has already been mentioned, is to
study the transonic properties and the stability issues of low angular momentum 
axisymmetric black hole accretion from a dynamical systems point of view. Flows 
around a non rotating black hole in three possible geometries, namely, the
constant height, the conical equilibrium and the hydrostatically balanced 
vertical equilibrium respectively, have been studied under the influence 
of a generalized pseudo-Schwarzschild black hole potential. Both polytropic
as well as the isothermal equation of state have been used to describe the
flow. The analytical formalism provided here ensures that even without performing 
explicit numerical integration along the streamlines, a reasonably clear 
picture of the behaviour of the phase trajectories of the flow can be 
well apprehended. This work, thus, can be considered as a crucial alternative 
approach to understand the behaviour of non-exactly solvable coupled differential 
equations describing autonomous dynamical systems, as they are to usually 
encountered while studying the
accretion processes around compact astrophysical objects. 

The stationary critical solutions, as well as the stability criteria of such 
stationary configurations have been analyzed. For stationary structure, one finds 
that the general 
profile of the parameter space marking the multi-critical flow solution remains 
unaltered for all three flow geometries, as well as for every black hole potential
available in the literature. The corresponding numerical domain of $\left[{\cal E},
\lambda,\gamma\right]$, for which the multi-criticality has been observed, is, however, 
different for different disc geometries, as well as for different potentials for 
a particular equation of state. 
\cite{bhyasan01} proposed that the three different disc models,
namely constant height flow, flow in conical equilibrium and in vertical
equilibrium should be identical, provided one can change the polytropic
constant in certain ways. In this way they suggested that a relativistic
flow of constant height might have same properties as an isothermal flow
in two other geometries. This is, however, not true.
One is able to argue that it is not possible to
allow accretion under various flow geometries identically by merely changing the 
polytropic index of the flow, since the shift from one flow geometry to another
is a non-continuous transformation in a multi-dimensional parameter space. 
Hence, it is certainly not a mathematically consistent approach to establish 
an equivalence between the polytropic flow in one type of disc geometry to 
the isothermal flow in the other by making a $\gamma$ transformation, since the 
corresponding first integrals of motion for the flow, derived on using the two 
different equations
of state (polytropic/adiabatic and isothermal) imply distinctively different 
physical properties.

The Mach number relation at the critical point differentiate the flow in vertical 
equilibrium ($M_c<1$ for polytropic flow) from the two other kinds of flow (constant
height flow and the flow under conical equilibrium), for which the critical 
point and the sonic point are identical (i.e., $M_c=1$), upon using 
both the equations of state. The result $M_c<1$ is a consequence of the 
effective speed of the propagation of the linear acoustic perturbation (sound 
speed) due to the presence of the polytropic sound speed $c_s$ in the 
expression of the disc height (which, in other words, is a direct 
repercussion of the vertical equilibrium assumption itself).

The stability properties of the aforesaid stationary configuration 
have been realized by perturbing (about the stationary configuration)
the full time-dependent flow equations in various disc geometries, 
under various black hole potentials, and then 
by observing the time evolution of such perturbations. The general form
of the wave equations corresponding to the dynamics of such 
perturbations, as well as the related acoustic metric, are identical for any 
black hole potential used, but is very different for different flow geometries. 
However, such perturbations do not diverge in any physical sense
for any kind of flow geometries. This ensures that the 
stability of the stationary configuration, at least for accretion around 
non-rotating black holes under various pseudo-Schwarzschild potentials,
is a generic feature independent of the nature of the space time (the 
explicit form of the black hole potential) as well as the geometric 
configuration of the flow (disc structure).

The exact forms of the corresponding acoustic metric for various disc
geometries have also been derived in this work. This will help in
studying the low angular momentum pseudo-Schwarzschild axisymmetric black hole
accretion as a natural example of analogue gravity phenomena. It has
recently been shown that the analogue surface gravity can be computed for
multi-critical accretion onto astrophysical black holes \citep{abd06,dbd07}.
The present work will allow to study such phenomena for various different
geometric configuration of the flow.

\section*{Acknowledgements}

This research has made use of NASA's Astrophysics Data System. 
SA would like to acknowledge the kind hospitality 
provided by HRI, Allahabad, India, under a visiting student
research programme. The visits of SN at HRI was partially supported 
by astrophysics project under the XIth plan at HRI.
The work of TKD has been partially supported by a research grant 
provided by ASIAA, Taiwan, under a guest scientist research 
programme. Discussion with Rukmini Dey regarding the properties 
of the Sylvester matrix is acknowledged.

%\bibliographystyle{mn2e}
%\bibliography{astro}
\end{document}